\documentclass[%
 reprint,
superscriptaddress,
nofootinbib,
nobibnotes,
 amsmath,amssymb,
 aps,
 prl,
]{revtex4-1}

\usepackage{siunitx}
\usepackage{graphicx}
\usepackage{dcolumn}
\usepackage{bm}
\usepackage{epstopdf}
\usepackage{bbm}
\usepackage{amsmath}

\usepackage[hidelinks]{hyperref}

\makeatletter
\newcommand{\mypm}{\mathbin{\mathpalette\@mypm\relax}}
\newcommand{\@mypm}[2]{\ooalign{%
		\raisebox{.1\height}{$#1+$}\cr
		\smash{\raisebox{-.6\height}{$#1-$}}\cr}}
\makeatother

\makeatletter
\newcommand*{\citenumns}[2][]{%
	\begingroup
	\let\NAT@mbox=\mbox
	\let\@cite\NAT@citenum
	\let\NAT@space\NAT@spacechar
	\let\NAT@super@kern\relax
	\renewcommand\NAT@open{}%
	\renewcommand\NAT@close{}%
	\cite[#1]{#2}%
	\endgroup
}
\makeatother

\usepackage{commath}
\usepackage{tabularx}

\usepackage{booktabs}

\begin{document}

\preprint{APS/123-QED}

\title{Intrinsic spin-orbit coupling gap and the evidence of a topological state in graphene}

\author{J. Sichau}
\thanks{These authors contributed equally}
\affiliation{%
	Center for Hybrid Nanostructures (CHyN), University of Hamburg, Luruper Chaussee 149, 22607 Hamburg, Germany\\
}
\author{M. Prada}%
\thanks{These authors contributed equally}
\affiliation{%
	I. Institute for Theoretical Physics, University of Hamburg, Jungiusstrasse 9-11, 20355 Hamburg, Germany\\
}
\author{T. J. Lyon}%
\affiliation{%
	Materials Science and Engineering, University of Wisconsin-Madison 1509 Engineering Drive, Madison, WI 53706, USA\\
}%
\author{B. Bosnjak}
\author{T. Anlauf}
\author{L. Tiemann}
\author{R.H. Blick}%
\affiliation{%
	Center for Hybrid Nanostructures (CHyN), University of Hamburg, Luruper Chaussee 149, 22607 Hamburg, Germany\\
}%

\date{\today}

\begin{abstract}
In 2005 Kane \& Mele 
[C. L. Kane and E. J. Mele, Phys. Rev. Lett. {\bf 95}, 226801 (2005)], 
predicted that at sufficiently low energy, 
graphene exhibits a topological state of matter with an energy gap 
generated by the atomic spin-orbit interaction. 
However, this intrinsic gap has not been measured to this date. 
In this letter, we exploit the chirality of the low energy states to 
resolve this gap. 
We probe the spin states experimentally, by employing low temperature
microwave excitation in a resistively detected electron spin resonance on graphene.
The structure of the topological bands is reflected in our transport experiments, 
where our numerical models allow us to identify the resonance signatures.
We determine the intrinsic spin-orbit bulk gap to be exactly $\SI{42.2}{\micro\electronvolt}$. 
Electron-spin resonance experiments can reveal the competition between
the intrinsic spin-orbit coupling and classical Zeeman energy that arises at low magnetic fields and demonstrate that graphene
remains to be a material with surprising properties.
\end{abstract}

\maketitle

In the early years of the rise of graphene, Kane \& Mele \cite{kane2005quantum,kane2005z} predicted that the symmetry-allowed spin orbit
potential in graphene gives rise to a spin-Hall insulating (SHI) state \cite{hasan2010colloquium}.
This novel electronic state of matter would be chiral and gapped in the bulk, while supporting spin transport along the
sample boundaries. 
The magnitude of the bulk gap, which is proportional to the atomistic or intrinsic spin-orbit coupling (SOC), 
determines the observability of an insulator phase of matter that is distinct from any ordinary insulator, 
characterized by chiral states. 
However, this intrinsic gap has not been experimentally established in graphene to this date and theoretical controversy
exists with regard to its precise magnitude \cite{min2006intrinsic,yao2007spin,konschuh2010tight}.

In this letter we  aim at resolving the intrinsic gap by coupling  mesoscopic Hall-bar  graphene 
structures at low temperatures to an external radio-frequency  source. 
We exploit the chirality of the low-energy bands and probe the distinct spin states experimentally, 
by employing microwave excitation in  resistively detected electron spin resonance (RD-ESR).
We detect two spectral lines of ESR as a function of magnetic field separated by a constant energy.  
An extended Dirac model allows us to identify this energy separation with  the  intrinsic SOC gap. 
 

In the Dirac model,  the notion of \textit{sublattice spin} is introduced, with `up' and `down' states 
being identified with the two
sublattice components, $u_A^K$ and $u_B^K$, respectively, that are centered around atoms of the A and B sublattices 
\cite{katsnelson2012graphene,neto2009electronic,hasan2010colloquium,geim2010rise}. 
In the bispinor basis ${\{\uparrow ,\downarrow\} \otimes \{u_A^K , u_B^K\}}$, 
the effective mass Hamiltonian near the Dirac points (DPs) $K$ and $K^\prime$ 
takes the form: 
\begin{equation}
H(\boldsymbol{k},\tau) = \hbar v_F\mathbb{I}_2\otimes(\tau\sigma_xk_x + \sigma_yk_y) + 
\lambda_I \tau_z s_z \otimes \sigma_z,
\label{eq1}
\end{equation}
where $\tau = \pm 1$ labels the valley $K$ ($K^\prime$), $\sigma_i$, $s_z$
are the Pauli matrices acting on the sublattice spin and real spin, respectively, 
$\boldsymbol{k}$ is the coordinate in reciprocal space with a 
DP at the origin, and $\mathbb{I}_2$ is the unitary $2\times 2$ matrix. 
The first term yields gapless states with the characteristic linear dispersion of massless Dirac fermions, 
$E(\boldsymbol{k}) = \pm v_F|\boldsymbol{k}|$. 
The degeneracy $\boldsymbol{k} = 0$ is protected by sublattice symmetry \cite{hasan2010colloquium}, 
and elsewhere, $\boldsymbol{\sigma}$ and $\boldsymbol{k}$ are \textit{collinear} and eigenstates of the 
Hermitian, unitary chirality operator $\hat{h}_k$ 
\cite{goerbig2011electronic,mikitik1999manifestation,suzuura2002crossover,katsnelson2006chiral,katsnelson2006zitterbewegung},
\begin{equation}
	\hat{h}_k = \hat{\boldsymbol{\sigma} \cdot \boldsymbol{k} / |\boldsymbol{k}|}.
	\label{eq:chirality}
\end{equation}
The chirality near $K$ is inverted with respect to the chirality around $K^\prime$ 
\cite{katsnelson2006chiral,katsnelson2006zitterbewegung,gorbar2002magnetic}. 
In essence, this dichotomy means that
an electron  state at $K$  and a hole state at $K^\prime$ 
are intricately connected by sublattice symmetry \cite{geim2010rise}. 
For samples with finite dimensions, this necessarily results in a topological phase, with the emergence 
of edge states connecting electron and hole bands at different DPs.
\begin{figure*}[]
	\centering
	\includegraphics[width=180mm]{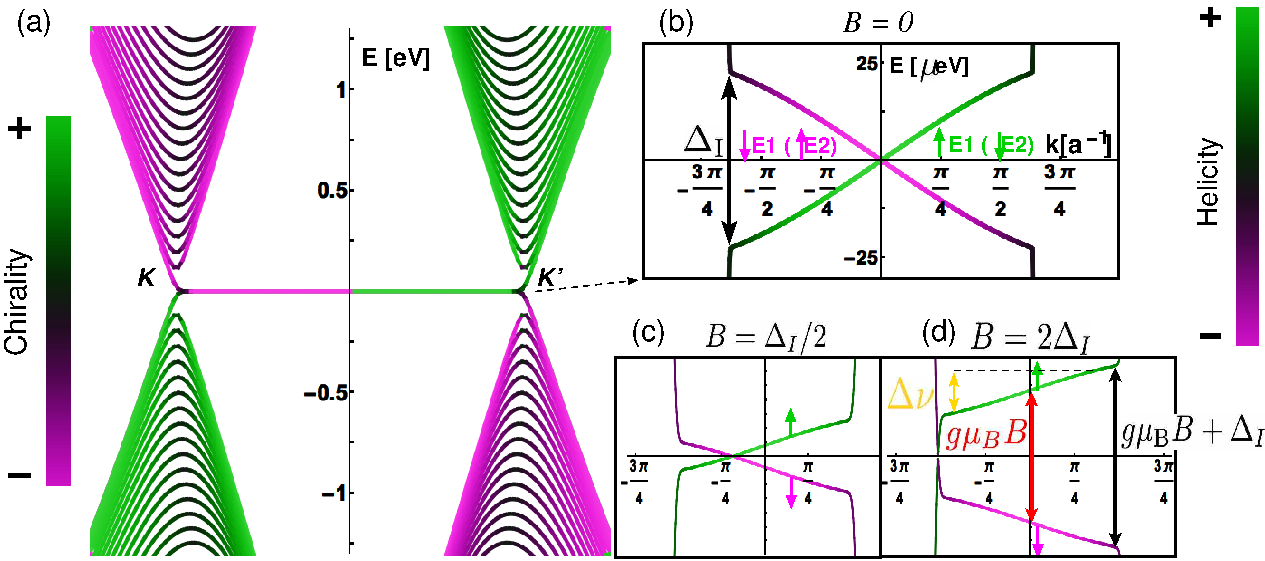}
	\caption[]{(a) Band structure of a honeycomb lattice terminated on a zig-zag edge with periodic boundary conditions along the armchair direction, colored according to their chirality, $\langle\hat{h}_k\rangle$: Green (magenta) denotes positive (negative) chirality. The states crossing the gap are flat on this energy scale (high DOS). 
		(b) Magnified dispersion relation near the Fermi level of the first Brillouin zone, showing the edge states in detail. 
		The bands are now colored according to their \textit{helicity},  $\langle\hat{h}_k\rangle_{\textrm{Edge}}$, 
		with black denoting ‘bulk’ character. 
		(c,d) Same as in (b) for the E1 bands, with $B= \Delta_I/2$ and $B= 2\Delta_I$, respectively, in units of $g\mu_{\rm B}$.
		The spin on each band is indicated with a color matching arrow. 
	}
	\label{fig:fig1}
\end{figure*}
The second term of Eq. (\ref{eq1}) is the effective intrinsic SOC \cite{kane2005quantum,kane2005z,hasan2010colloquium}, 
and is mostly originated from the poorly occupied d-orbitals \cite{gmitra2009band}. This term
respects sublattice, parity and time-reversal symmetries, and opens up a bulk gap of opposite sign at 
each DP of magnitude $|\Delta_I| = 2\lambda_I$. 
The energetically low-lying edge eigenstates become locally \textit{helical}, with collinear spin $\boldsymbol s$ and sublattice 
spin $\boldsymbol{\sigma}$, 
as illustrated in Fig. \ref{fig:fig1}c. 
The projection of the \textit{chirality} $\hat{h}$ onto the edges is then isomorphic with the Hermitian, unitary 
\textit{helicity} operator 
$\langle \hat{h}_k \rangle_{\textrm{Edge}} = 2 \langle \hat{\sigma}_z \hat{s}_z \rangle_{\textrm{Edge}}$, where 
$\langle...\rangle_{\textrm{Edge}}$ means that the evaluation is obtained by projecting onto the edge’s local 
density of states (LDOS). 
In practical terms, this means that \textit{the midgap states are spin and sublattice-spin polarized}, 
with the corresponding pseudovectors being either parallel or anti-parallel.

Fig. \ref{fig:fig1}a represents the dispersion of a graphene slab (see supplemental material), where the bands
are colored according to their chirality $\langle\hat{h}_k\rangle$. 
In Fig. \ref{fig:fig1}b, we zoom into the low-lying energy states $E \sim |\Delta_I|$. 
In order to distinguish the edges from the bulk bands, we color the bands according to their helicity 
$\langle\hat{h}_k\rangle_{\textrm{Edge}}$, with black denoting now bulk-like bands. 
The bulk shows a gap of $\Delta_I = \pm 2 \lambda_I$ of opposite sign at either DP, as expected \cite{novoselov2005two}, 
while the edges are ungapped. 
Those midgap bands are doubly-degenerate pseudo-spin pairs, with $\sigma_z = \pm 1$, located at either edge. 
At edge E1, the spin ‘up’ states have indeed positive velocity, $\partial E/\partial k = v_F^{\textrm{Edge}} > 0$ 
(green, Fig. \ref{fig:fig1}b), whereas those with spin ‘down’ travel backwards 
$\partial E/\partial k = -v_F^{\textrm{Edge}} < 0$ (magenta, Fig. \ref{fig:fig1}b). 
The converse occurs in E2, with spin ‘up’ (‘down’) showing positive (negative) velocity, that is, 
\textit{E2 is related to E1 by a mirror reflection}.

When a magnetic field $B$ is applied perpendicularly to the graphene sheet, the Kramers’ pairs split into spin-up 
and -down levels by the Zeeman energy, $g\mu_{\rm B}B$. In Fig. \ref{fig:fig1}c and \ref{fig:fig1}d  we plot, for clarity, 
only the midgap E1 bands, and color them again according to the LDOS. 
The edge's occupation is maximal at the ${\Gamma}$ point, spreading over a bandwidth given by $h\Delta\nu$ 
(gold double headed arrows of Fig. \ref{fig:fig1}d). 
When the Zeeman energy is below the SOC gap,  
opposite spin band-crossing pairs occur at the Fermi level, and are predominantly localized at one edge (Fig. \ref{fig:fig1}b). 
The SHI phase is preserved, that is: for the $k$-interval $[0,\pi]$, we encounter an edge state that crosses the Fermi 
level once, for each spin sector. At $g\mu_{\rm B}|B|>|\Delta_I|$, the SHI is no longer preserved, as the bands at the 
crossings have bulk character. 
A gap  centered at the Zeeman energy opens between the opposite spin edge bands (red arrow of Fig. \ref{fig:fig1}d).

\begin{figure}[]
	\centering
	\includegraphics[width=.35\textwidth]{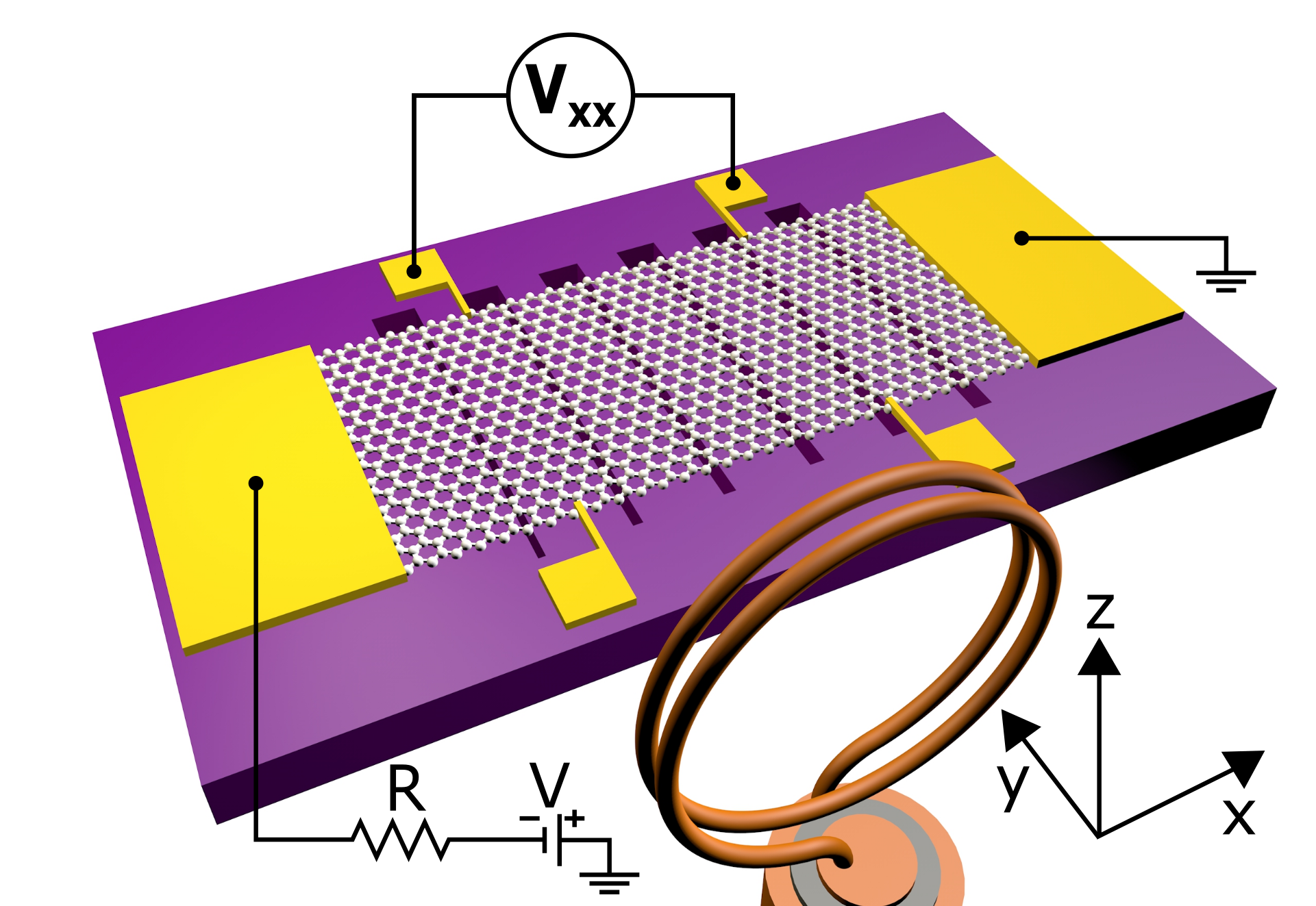}
	\caption[]{Schematics of the measurement setup with monolayer graphene patterned into a Hall bar structure. 
A nearby loop antenna excites the system (not to scale).}
	\label{fig:fig2}
\end{figure}

\begin{figure}[]
	\centering
	\includegraphics[width=86mm]{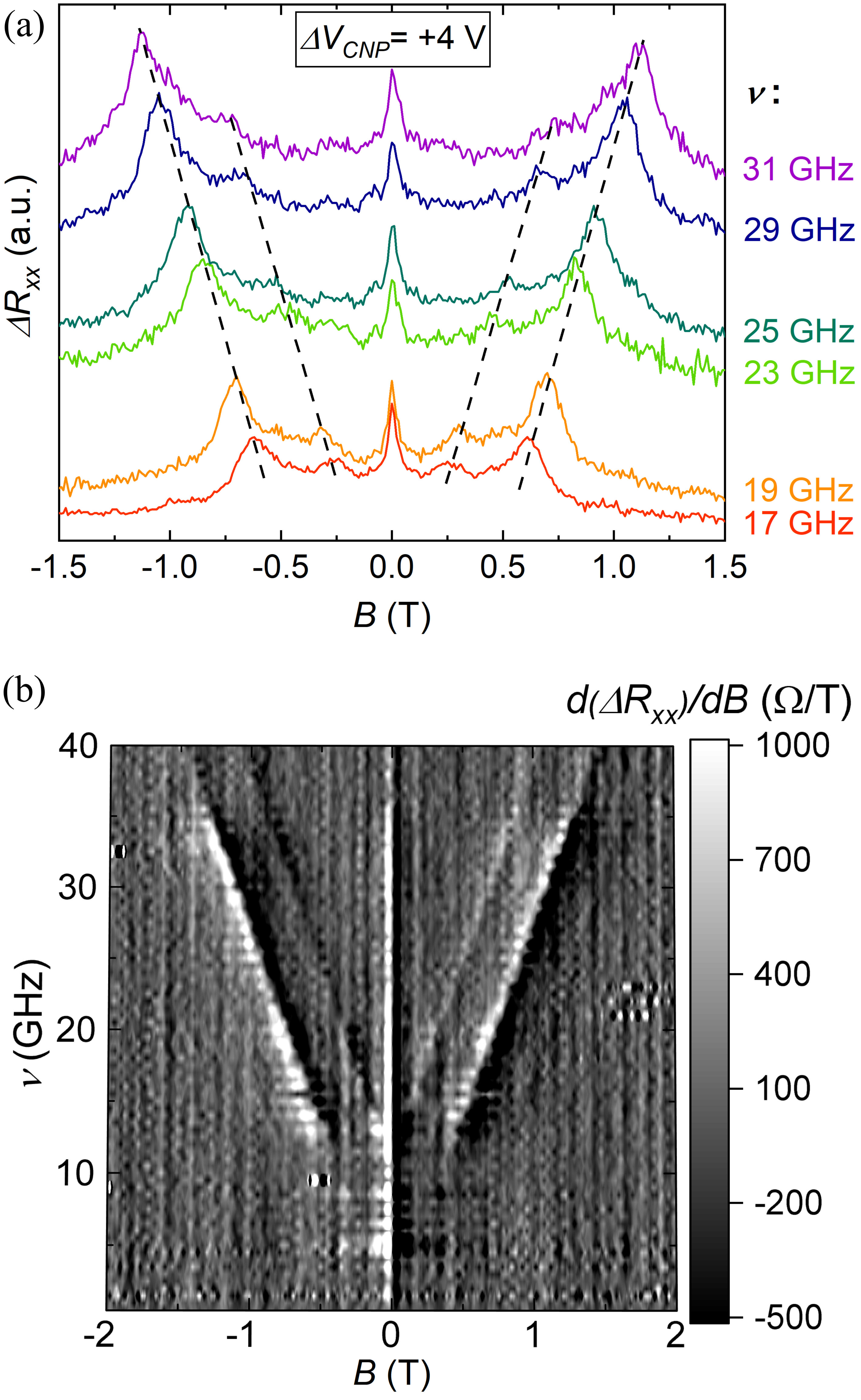}
	\caption[]{(a) Individual measurements at $T = \SI{4.2}{\kelvin}$ for various frequencies with constant gate voltage $\Delta V_{\rm {CNP}} = V_g-V_{\rm {CNP}} = \SI{4}{\volt}$. Two resonances which are symmetric in $B$ exhibit a linear dependence on $\nu$ (dashed lines). The feature at zero field stems from the the weak localization in the sample. The data has been shifted and scaled for clarity. (b) Derivative data of all measurements recorded for $\SI{0.5}{\giga\hertz} \leq \nu \leq \SI{40}{\giga\hertz}$. The resonance signal is present over a wide range of frequencies except for $\nu\lesssim \SI{11}{\giga\hertz}$. The upper "V"-feature intercepts the frequency axis at $\nu_1 = \SI{10.2}{\giga\hertz}$, while the lower feature interpolates to $\nu_2 = 0$. This difference corresponds to an energy of $\Delta E = \SI{42.2}{\micro\electronvolt}$.
		}
	\label{fig:fig3}
\end{figure}

We address these opposite-spin, helical  edge bands by employing RD-ESR \cite{mani2012observation,lyon2017probing}, 
a spin-sensitive probing technique that  couples carriers of opposite spin by microwave excitation, 
and detects the response resistively.  
Our ESR measurements are performed on a Hall-bar graphene structure of \SI{200}{\micro\meter} 
length and \SI{22}{\micro\meter} width with an intrinsic charge carrier 
density and mobility of \SI{2e11}{\centi\meter^{-2}} and \SI{3760}{\centi\meter^2\volt^{-1}\second^{-1}}, 
respectively \cite{lyon2017upscaling}. 
We minimize the unwanted external SOC sources \cite{ rashba2009graphene,platzman1973waves,konschuh2010tight} 
and the effective contact area of the graphene with the substrate \cite{lyon2017probing} 
by suspending the graphene sheet on a trenched SiO$_2$ layer at zero gate voltage (see supplementary information).
Microwave excitation is applied through a loop antenna next to the sample (see Fig. \ref{fig:fig2}). 
The longitudinal sample resistance, $R_{xx}$, is then probed as a function of the magnetic field $B$, 
both in the absence ($R_{xx,\textrm{dark}}$) and in the presence ($R_{xx,\nu}$) of microwave radiation. 
Illuminating the sample reduces the overall resistance, as more conducting bands become populated. 
Moreover, a signal in the photo-induced differential resistance, 
$\Delta R_{xx}(\nu) = R_{xx,\textrm{dark}}-R_{xx,\nu}$ is expected whenever the carrier Zeeman 
splitting matches the microwave energy of the bulk, $h\nu = (2\lambda_I \pm g\mu_bB)$ or that of 
the edges, $h\nu = \pm g\mu_{\rm B}B$, as dictated by spin selection rules  
(black and red double-headed arrows of Fig. 1d, respectively). 
At these matching frequencies, the band population increases
and the resistance is consequently reduced, revealing a peak in $\Delta R_{xx}(\nu)$ \cite{lyon2017probing}. 
We emphasize that unlike ideal infinite graphene, a finite DOS exists near the charge neutrality point that originates from 
edge states (note their rather flat dispersion in Fig. \ref{fig:fig1}).  
Due to unintentional doping, the Fermi energy is then only shifted  by $\Delta E_F\simeq 0.1$meV  
(see supplemental material), an amount comparable to $k_BT$, and 
thus allowing a finite amount of thermally excited carriers even within the gap, 
$f(\Delta_I) = (1+e^{\Delta E_F/k_BT})^{-1} \gtrsim 0.06$. 
On the other hand, 
the 
Maxwell-Boltzmann distribution dictates $n_\uparrow/n_\downarrow = e^{-g\mu_B B/k_BT} \simeq 0.9$ for 
$g\mu_{\rm B} |B|\simeq \Delta_I$, 
allowing for a detectable signal by net energy absorption even at energies comparable to the intrinsic gap \cite{Slichter1990}. \\
In Fig. \ref{fig:fig3}(a), $\Delta R_{xx}(\nu)$ is plotted for multiple frequencies, exhibiting a linear dependence of 
the resonance frequency in magnetic field. Fig. \ref{fig:fig3}(b) shows the derivative of $\Delta R_{xx}(\nu)$ in the frequency-magnetic field plane. The two salient "V"-shaped features are separated by a constant frequency of 
$\nu \approx \SI{10.2}{\giga\hertz}$ (\SI{42.2}{\micro\electronvolt}). 

The extrapolation of the prominent lower feature intersects with the axis at its origin, representing the 
edge's Zeeman splitting. When the Zeeman splitting is smaller than the intrinsic gap, the edge states cover the entire 
range of energies within $\Delta_I$ (Fig. \ref{fig:fig1}c, \ref{fig:fig1}d), as the bands of opposite spin cross at the 
Fermi energy.  
However, as the Zeeman splitting overcomes the intrinsic gap, a band of forbidden energies opens up for the edges, 
allowing an ESR signal to be detected (solid red arrows in Fig. 1d). 
This is reflected in the strong signal for $|B| \gtrsim |\Delta_I|/g\mu_{\rm B}$ and in the absence of signal, otherwise. 
The large intensity of the signal is related to the large DOS of the edges and its width $\Delta\nu$ is 
related to their dispersive character (see Fig. \ref{fig:fig1}d). 

We identify the upper "V"-feature with the bulk signal, $h\nu = (\Delta_I \pm g\mu_{\rm B}B)$ (black double headed arrows 
of Fig. \ref{fig:fig1}c). It reveals a zero-field splitting, which is a direct measurement of the \textit{intrinsic} 
SOC splitting $\Delta_I$: $\nu(B=0) = \SI[separate-uncertainty=true]{10.2\pm 0.2}{\giga\hertz}$, and in energy
$\Delta E = \SI[separate-uncertainty=true]{42.2\pm0.8}{\micro\electronvolt}$. 
Its weaker intensity reflects the lower DOS of the \textit{bulk}. 
\begin{figure}[]
	\centering
	\includegraphics[width=86mm]{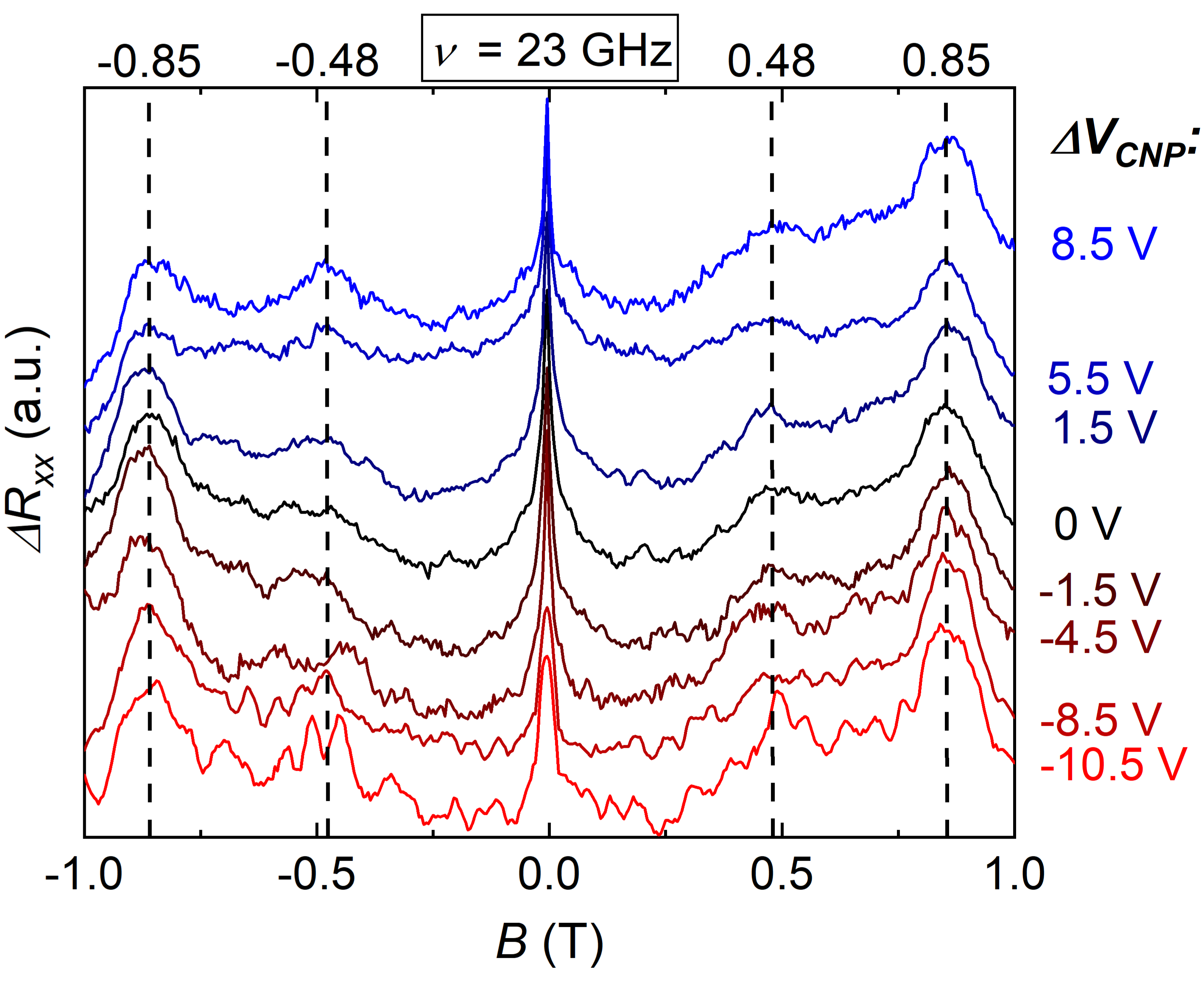}
	\caption[]{Individual measurements at $T = \SI{1.4}{\kelvin}$ for various gate voltages $\Delta V_{\rm {CNP}}$ with constant frequency $\nu = \SI{23}{\giga\hertz}$, with red (blue) indicating electron (hole) character of the main charge carriers. The two symmetric resonance peaks are evident at $|B_1| = \SI{0.48}{\tesla}$ and $|B_2| = \SI{0.85}{\tesla}$ (dashed lines), and are invariant to gate voltage, \textit{i.e.}, to charge carrier density and type. The data has been shifted and scaled for clarity.}
	\label{fig:fig4}
\end{figure}
Moreover, our value is consistent with a zero-field splitting of \SI{10.76}{\giga\hertz} reported by 
Mani \textit{et al.} \cite{mani2012observation} on three small, epitaxially grown graphene samples on SiC substrate. 
The authors did not identify the intrinsic gap to be responsible for their observations, however, their coincident results
strongly support our claim: the zero-field splitting  corresponds to an 
{\it intrinsic} property of graphene, namely, its intrinsic SOC gap, which makes it a sample independent effect. \\
These measurements have been reproduced under different conditions of temperature and carrier densities and in different samples. 
Fig. \ref{fig:fig4} shows the data for a $\SI{1}{\milli\meter} \times \SI{100}{\micro\meter}$ graphene Hall bar on a flat SiO$_2$ 
substrate at a temperature of \SI{1.4}{\kelvin}. 
The two pairs of resonances occur at the same magnetic fields as for the sample of Fig. \ref{fig:fig3}, 
and they are found to be invariant over a wide range of gate voltages. This excludes other possible zero-field splitting 
candidates, as \textit{e.g.} Rashba 
$H_R = \lambda_R (\boldsymbol{s}\otimes\boldsymbol{\sigma})\hat{z}$ \cite{gmitra2009band,rashba2009graphene}. We note that 
including $H_R$ leaves indeed the SHI picture invariant as long as 
$\lambda_R<\lambda_I$ \cite{kane2005quantum,kane2005z} (see also supplemental material). 
Finally, we note that for the large sample dimensions 
we consider in this work, we can safely assume that the edge and bulk signals are width insensitive:  
Localized solutions for other edge types, such as armchair or ragged
edges, yield qualitatively similar results due to the bulk-edge correspondence and the continuum limit
\cite{winkler2017effective}.

The magnitude of the intrinsic gap in graphene determines the observability of the SHI phase, but 
has been subject of theoretical controversy: After its initial rough estimate of about \SI{100}{\micro\electronvolt} by 
Kane \textit{et al.} \cite{kane2005z}, Min \textit{et al.} \cite{min2006intrinsic} and Yao \textit{et al.} \cite{yao2007spin} 
reported independently a theoretical calculation of \SI{1}{\micro\electronvolt}. 
Konschuh \textit{et al.} \cite{konschuh2010tight} and Boettger \textit{et al.} \cite{boettger2007first} 
used first-principles calculations to deliver a larger value, around the $25-\SI{50}{\micro\electronvolt}$. 
Our experimental measurement agrees best with this range, rendering the SHI experimentally accessible for graphene.

In graphene, the symmetry protected sublattice degeneracy favors the emergence of a fascinating state of matter, the SHI. 
We find its presence encoded in exotic transitions that can be observed in RD-ESR experiments. 
To illuminate the origin of these ESR transitions and the underlying complex band structure in 
suspended graphene, we have extended the conventional Dirac model and characterized the bands 
according to their relevant quantum numbers and properties. 
The existence of helical carriers with a linear dispersion offers a testbed for 
the studies of the fundamental massless Dirac fermions and anti-fermions.\\
\newline
This work has been supported by the excellence cluster ’The Hamburg Centre for Ultrafast Imaging - Structure, Dynamics and Control of Matter at the Atomic Scale’ of the Deutsche Forschungsgemeinschaft (EXC-1024). We thank M. I. Katsnelson, D. Pfannkuche, and A. Lichtenstein for fruitful discussions. R.H.B. likes to thank Klaus von Klitzing for discussion. \\
\newline

\bibliography{bibliography}

\end{document}